\documentclass[letter,prd,aps,twocolumn,floatfix,superscriptaddress]{revtex4}

\usepackage{amssymb,amsmath,graphicx}
\usepackage{hyperref}
\usepackage[usenames,dvipsnames]{color}
\usepackage{array}
\makeatletter
\renewcommand*{\p@section}{\S\,}
\renewcommand*{\p@subsection}{\S\,}

\makeatother

\def\jnl@style{\it}
\def\aaref@jnl#1{{\jnl@style#1}}

\def\aaref@jnl#1{{\jnl@style#1}}

\def\aj{\aaref@jnl{AJ}}                   
\def\araa{\aaref@jnl{ARA\&A}}             
\def\apj{\aaref@jnl{ApJ}}                 
\def\apjl{\aaref@jnl{ApJ}}                
\def\apjs{\aaref@jnl{ApJS}}               
\def\ao{\aaref@jnl{Appl.~Opt.}}           
\def\apss{\aaref@jnl{Ap\&SS}}             
\def\aap{\aaref@jnl{A\&A}}                
\def\aapr{\aaref@jnl{A\&A~Rev.}}          
\def\aaps{\aaref@jnl{A\&AS}}              
\def\azh{\aaref@jnl{AZh}}                 
\def\baas{\aaref@jnl{BAAS}}               
\def\jrasc{\aaref@jnl{JRASC}}             
\def\memras{\aaref@jnl{MmRAS}}            
\def\mnras{\aaref@jnl{MNRAS}}             
\def\pra{\aaref@jnl{Phys.~Rev.~A}}        
\def\prb{\aaref@jnl{Phys.~Rev.~B}}        
\def\prc{\aaref@jnl{Phys.~Rev.~C}}        
\def\prd{\aaref@jnl{Phys.~Rev.~D}}        
\def\pre{\aaref@jnl{Phys.~Rev.~E}}        
\def\prl{\aaref@jnl{Phys.~Rev.~Lett.}}    
\def\pasp{\aaref@jnl{PASP}}               
\def\pasj{\aaref@jnl{PASJ}}               
\def\qjras{\aaref@jnl{QJRAS}}             
\def\skytel{\aaref@jnl{S\&T}}             
\def\solphys{\aaref@jnl{Sol.~Phys.}}      
\def\sovast{\aaref@jnl{Soviet~Ast.}}      
\def\ssr{\aaref@jnl{Space~Sci.~Rev.}}     
\def\zap{\aaref@jnl{ZAp}}                 
\def\nat{\aaref@jnl{Nature}}              
\def\iaucirc{\aaref@jnl{IAU~Circ.}}       
\def\aplett{\aaref@jnl{Astrophys.~Lett.}} 
\def\apspr{\aaref@jnl{Astrophys.~Space~Phys.~Res.}}
\def\bain{\aaref@jnl{Bull.~Astron.~Inst.~Netherlands}} 
\def\fcp{\aaref@jnl{Fund.~Cosmic~Phys.}}  
\def\gca{\aaref@jnl{Geochim.~Cosmochim.~Acta}}   
\def\grl{\aaref@jnl{Geophys.~Res.~Lett.}} 
\def\jcp{\aaref@jnl{J.~Chem.~Phys.}}      
\def\jgr{\aaref@jnl{J.~Geophys.~Res.}}    
\def\jqsrt{\aaref@jnl{J.~Quant.~Spec.~Radiat.~Transf.}}
\def\memsai{\aaref@jnl{Mem.~Soc.~Astron.~Italiana}}
\def\nphysa{\aaref@jnl{Nucl.~Phys.~A}}   
\def\physrep{\aaref@jnl{Phys.~Rep.}}   
\def\physscr{\aaref@jnl{Phys.~Scr}}   
\def\planss{\aaref@jnl{Planet.~Space~Sci.}}   
\def\procspie{\aaref@jnl{Proc.~SPIE}}   

\begin{document}

\date{\today}
\title{The $m=1$ instability \& gravitational wave signal in 
binary neutron star mergers}

\author{Luis Lehner}
\affiliation{Perimeter Institute for Theoretical Physics,Waterloo, Ontario N2L 2
Y5, Canada}
\author{Steven L. Liebling}
\affiliation{Department of Physics, Long Island University, Brookville, New York
 11548, USA}
\author{Carlos Palenzuela}
\affiliation{Departament de F\'isica, Universitat de les Illes Balears and Insti
tut d'Estudis Espacials e Catalunya, Palma de Mallorca, Baleares E-07122, Spain}
\author{Patrick M.  Motl}
\affiliation{School of Sciences, Indiana University Kokomo, 
Kokomo, IN 46904, USA}


\begin{abstract}
We examine the development and detectability 
of the $m=1$ instability in the remnant of binary neutron star mergers. 
The detection of the gravitational mode associated with the $m=1$ degree of freedom
could potentially reveal details of the equation of state.
We analyze the post-merger epoch of simulations of both equal and non-equal 
mass neutron star mergers using three realistic, microphysical equations of state
and neutrino cooling. Our studies
show such an instability develops generically and within a short dynamical time to strengths
that are comparable or stronger than the $m=2$ mode which is the strongest during the 
early post-merger stage. We estimate the signal to noise
ratio that might be obtained for the $m=1$ mode and discuss the prospects
for observing this signal with available Earth-based detectors. 
Because the $m=1$ occurs at roughly half the frequency of the more powerful 
$m=2$ signal and because it can potentially be long-lived,
targeted searches could be devised to observe it. 
%
%
We estimate that with constant amplitude direct detection of the mode
could occur up to a distance of roughly $14\,\mathrm{Mpc}$
whereas a search triggered by the inspiral signal could extend this distance to roughly $100\,\mathrm{Mpc}$.
\end{abstract}

\maketitle

\tableofcontents

\section{Introduction}\label{introduction}
The era of gravitational wave astronomy has begun with the
spectacular detection of gravitational waves from  event GW150914~\cite{Abbott:2016blz}.
This detection not only established that advanced interferometers can indeed detect the
tiny effect of gravitational waves reaching the detectors, but it also demonstrated that data analysis using
numerical relativity can extract physical parameters of the underlying engine. Assuredly the
various efforts on the experimental hardware, data analysis, and source modeling are now
being pursued with even more ardor.

Although a black hole binary produced the first direct detection of gravitational waves,
the merger of two neutron stars had long been a primary target for Advanced LIGO~(aLIGO), and indeed a detection
of a binary neutron star system is eagerly anticipated. One reason for this excitement
is that neutron stars represent a very extreme state of matter not accessible in the laboratory,
and gravitational waves may reveal the equation of state~(EoS) of matter at such extreme densities.

To understand what these gravitational waves tell us about the equation of state
requires modeling the merger of these neutron stars, and such an effort spans
a few techniques. 
A Post-Newtonian expansion of the two body problem is appropriate for the early
orbiting stage of the merger while numerical relativity is needed for
the near-merger, merger and post-merger phases~\cite{Faber:2012rw}.
Near coalescence (prior to the merger), numerical relativity is being used to enhance perturbative 
approaches (e.g. the effective one-body approach) in order to account for
tidal effects---that depend on the EoS---in a concise and 
analytic manner~\cite{Hinderer:2016eia}. 

The effects of the EoS become most significant as the stars approach each other
and tidal forces grow. However, this regime is characterized by frequencies higher
than those of the inspiral, and current detectors lose sensitivity for these increasing
frequencies.
Therefore,
the extraction of the EoS from gravitational waves alone can be a subtle and
difficult enterprise (e.g.~\cite{Lackey:2014fwa,Agathos:2015uaa}).

Consequently, exploring the various pathways to extract physical 
information during the coalescence and post-merger stages is critical.
These pathways include
the search for gravitational wave features highly sensitive to the EoS. In addition,
combining the complementary information provided by possible
electromagnetic counterparts~\cite{Lehner:2016lxy,Sekiguchi:2015dma,Rosswog:2015nja}
may greatly increase the information we can extract from the gravitational wave signal alone.

One example of a gravitational wave feature that may lead to better understanding of
the EoS concerns the frequency spectrum of the post-merger $l=2,m=2$ mode. Several studies
have found that this post-merger signal is characterized by a peak frequency that
is related to the underlying EoS~\cite{2015PhRvD..92d4045P,2015PhRvD..91f4027K,Foucart:2015gaa,Lehner:2016lxy}.

Beyond this dominant mode, recent work has also shown that a weaker but longer lived
mode, the $l=2,m=1$ mode, develops in the post-merger epoch of binary neutron star mergers.
This possibility was first described in Refs.~\cite{Ou:2006yd,Corvino:2010yj} using Newtonian gravity,
and more recently the growth of this mode has been found within full, general relativity 
simulations~\cite{2008PhRvL.100s1101A,East:2015vix,Dietrich:2015pxa,Radice:2016gym}.

Although initially weaker than the $l=2,m=2$, the $l=2,m=1$ has a couple compensating effects working for it.
First, the $m=1$ mode occurs at roughly half the frequency of the dominant mode, and at this
lower frequency the noise level is reduced with respect to that at higher frequencies\footnote{
At frequencies beyond $\approx$kHz the detector noise increases quadratically with frequency.}.
Second, because the $m=1$ mode is less radiative and because it has hydrodynamic and magnetic instabilities helping
support it, the mode is generally less damped than the $m=2$ mode,
and its longer life yields more signal to detect.
Furthermore, both these characteristic $m=1$ and $m=2$ modes are amenable to searches {\em triggered}
by the strong inspiral detection.
Moreover, the {\tt 2:1} frequency ratio of these two modes allows for an analysis of
the correlation between the frequencies that may improve detectability.

In this note, we go extend the studies provided in~\cite{Radice:2016gym,East:2015vix}  on
the development and detectability prospects of the $m=1$ mode.
In particular we illustrate the behavior of this mode in the mergers of both equal and 
unequal mass binaries with different realistic, microphysical equations of state. 
We examine the effects of neutrino cooling and changes to the EoS, and find
that even when accounting for the cooling due to neutrinos --which take away significant
from the resulting object-- the $m=1$ mode develops and grows to relevant strengths to impact
the characteristics of the gravitational waves emitted by the system. 
By varying the  mass ratio, we estimate the strength of this $m=1$ signal, and conclude that, as one would expect, the $m=1$ is much stronger for the unequal mass cases in the early stages after-merger.
We also estimate the expected SNR and discuss detectability prospects. In particular
we argue that such mode can be targeted following a triggered-search strategy.
We summarize details of our implementation in section~II, 
present results in section~III, and conclude in section~IV.

\section{Numerical Implementation}
\label{sec:implementation}

The evolution equations for the spacetime and the fluid are described in our previous paper~\cite{2015PhRvD..92d4045P},
while full details of our implementation are described in Ref.~\cite{Neilsen:2014hha}. Unless
otherwise specified, we adopt geometrized units  where $G=c=M_\odot = 1$, except for
some particular results which are reported more naturally in physical cgs units.

Einstein equations in the presence of both matter and radiation are,
\begin{equation}
G_{ab} = 8\,\pi\, (T_{ab} + {\cal R}_{ab}) \, ,
\end{equation}
where $T_{ab}$ is the stress energy tensor of a perfect fluid and ${\cal R}_{ab}$ is
the contribution from the radiation field.
Our stars consist of a fluid described by the stress energy tensor
\begin{equation}
T_{ab} = h u_a u_b + P g_{ab}  \, ,
\end{equation}
where $h$ is the fluid's {\it total} enthalpy $h \equiv  \rho (1+\epsilon) + P$,
and $\{\rho,\epsilon,Y_e,u^a,P\}$ are the rest mass energy density,
specific internal energy, electron fraction (describing the relative abundance of electrons compared to the total number of baryons), four-velocity, and pressure of the fluid,
respectively.  

To track the composition of the fluid and the emission of neutrinos we employ a leakage scheme.
In particular, the equations of motion consist of the following conservation laws
\begin{align}
\nabla_a T^a_b &= {\cal G}_b\, , \label{eq:DT}\\
\nabla^a (T_{ab} n^b) &= 0\, , \label{eq:DTN} \\
\nabla_a (Y_e \rho u^a) &= \rho R_Y\, ,\label{eq:divYe}
\end{align}
where $n^a$ is a timelike vector orthonormal to constant time surfaces employed
to express Einstein equations as an initial value problem (e.g.~\cite{Lehner:2001wq}). The sources ${\cal
  G}_a$~($\equiv -\nabla_c {\cal R}^c_{a}$) and $R_Y$ correspond to the radiation
four-force density and lepton source, respectively, and these quantities are computed within the
leakage scheme. 
These equations are conservation laws for the stress-energy tensor, 
matter, and lepton number, respectively.
 Notice that, in the absence of lepton source terms ($R_Y=0$),
Eq.~(\ref{eq:divYe}) provides a conservation law for leptons and is similar to
the familiar baryon conservation law, i.e., $Y_e$ is a mass scalar.

As mentioned, full details of our implementation can be found in Ref.~\cite{Neilsen:2014hha}, but a summary of our numerical techniques
is included for completeness. We employ finite difference techniques on a regular, Cartesian grid to discretize the system.
The geometric fields are discretized
with a fourth order accurate scheme that satisfies summation by
parts~\cite{SBP2,SBP3}, while a High-Resolution Shock-Capturing method based on the HLLE
flux formulae with PPM reconstruction is used to discretize the hydrodynamical variables. 
The fluid equations are discretized with finite
differences (rather than finite volume) as prescribed for the
third-order ENO method~\cite{ShuOsherI,Anderson:2006ay}. This simplifies
coupling the fluid equations to the Einstein equations.
The time evolution of the resulting equations adopts a third order accurate 
Runge-Kutta scheme~\cite{Anderson:2006ay,Anderson:2007kz}.
To provide sufficient resolution efficiently,                      we employ
adaptive mesh refinement~(AMR) via the HAD computational infrastructure. This infrastructure 
provides distributed, Berger-Oliger style AMR~\cite{had_webpage,Liebling} with
full sub-cycling in time, as well as an improved treatment of artificial boundaries~\cite{Lehner:2005vc}.

Importantly, we study these systems in 3D without imposing any
symmetry condition that might suppress certain dynamics.
For example, 
enforcing reflection symmetry in the equal mass, non-spinning case would necessarily exclude 
any odd $m$ mode.

\section{Results: Density behavior \& gravitational waves}
\label{sec:numericalresults}
This note studies the post-merger stage of binary neutron star systems 
that have been the focus of our recent work~\cite{2015PhRvD..92d4045P,Lehner:2016lxy}.
We extend these studies with the goal of examining more closely the development 
of an $m=1$ mode as a result of the merger of both equal and unequal mass cases.
 
The systems we study are
consistent with current astrophysical observations 
(see e.g.  Ref.~\cite{doi:10.1146/annurev-nucl-102711-095018}). 
Here, we consider binaries with the same total 
gravitational mass $M=2.70 M_{\odot}$ but with different mass ratios,
$q \equiv M_1/M_2$, ranging from  $q=1$ (the equal mass case) to $q=0.76$. 

We concentrate on the three realistic EoS previously 
considered \footnote{These three EoS are capable of producing neutron stars with masses of 
at least 2\,$M_\odot$, and they are 
thus consistent with current observations of NS masses~\cite{Demorest:2010bx,Antoniadis:2013pzd}.} in Refs.~\cite{2015PhRvD..92d4045P,Lehner:2016lxy}. These three EoS span a range of stiffnesses, from the
softest (smallest neutron stars) SFHo~\cite{2013ApJ...774...17S},  to the intermediate DD2~\cite{2012ApJ...748...70H},  and finally to the stiffest NL3~\cite{2012ApJ...748...70H}.

The physical parameters of the binaries 
and of our grid setup are summarized in Table~\ref{table:equal_mass}.
Notice that for the total mass considered here, the hypermassive neutron star~(HMNS) resulting 
from the SFHo merger collapses to a black hole roughly $\approx 8$ms after
merger. Thus the associated SNR for the gravitational waves emitted after merger for this case 
will likely be lower than the other cases.

\begin{table*}[t]\centering
\begin{ruledtabular}
\begin{tabular}{lllllllllllll}
EoS & q & $\nu$ 
 &  $m_{b}^{(1)}, m_{g}^{(1)}$ &  $m_{b}^{(2)}, m_{g}^{(2)}$  
  & $R^{(1)}$  & $R^{(2)}$ 
  & $C^{(1)}$  & $C^{(2)}$  
 & $\Omega_0$ \\ 

 &  &  & ~~[$M_{\odot}$] &  ~~[$M_{\odot}$]  
 & [km] & [km] &  &   
 & [rad/s]  \\

\hline
 NL3  & 0.85  & 0.248 & 1.34, 1.25 & 1.60, 1.47 & 14.75 & 14.8 & 0.125 & 0.147  & 1777 
  \\
 DD2  & 1.0  & 0.250 & 1.49, 1.36 & 1.49, 1.36 & 13.22 & 13.22 & 0.152 & 0.152  & 1776 
  \\ 
 DD2  & 0.85  & 0.248 & 1.36, 1.29 & 1.62, 1.47 & 13.20 & 13.25 & 0.144 & 0.164  & 1775 
  \\ 
 DD2  & 0.76 & 0.245 & 1.27, 1.18 & 1.71, 1.54 & 13.16 & 13.25 & 0.132 & 0.172   & 1775
  \\ 
 SFHo  & 0.85 & 0.248 & 1.37, 1.25 & 1.63, 1.47 & 11.95 & 11.85 & 0.154 & 0.183  & 1773 
  \\ 

\end{tabular}
\end{ruledtabular}
\caption{Summary of the binary neutron star systems considered in
this work. The initial data were computed using the {\sc Bin star}
solver from the {\sc Lorene} package~\cite{lorene}, with the
assumption that the stars have an initial constant temperature of
$T=0.02$~MeV and are in beta-equilibrium. All the binaries have a
total mass $M_{0}^{\rm ADM} = 2.7 M_{\odot}$ and start from an
initial separation of $45$~km. The outer boundary is located at
$750$~km and the highest resolution covering both stars is $\Delta
x_{\rm min}=230 m$. The table displays the mass ratio of the binary
$q \equiv M_1/M_2$ and $\nu=M_1 M_2/M^2 = q/(q+1)^2$, the baryon (gravitational) mass
of each star $m_b^{(i)}$ ($m_g^{(i)}$), its circumferential radius $R^{(i)}$ and
its compactness $C^{(i)}$ (i.e., when the stars are at infinite
separation),  the initial
orbital angular frequency $\Omega_0$.
}
\label{table:equal_mass}
\end{table*}

\begin{figure}[h]
\centering
\includegraphics[width=8.5cm,angle=0]{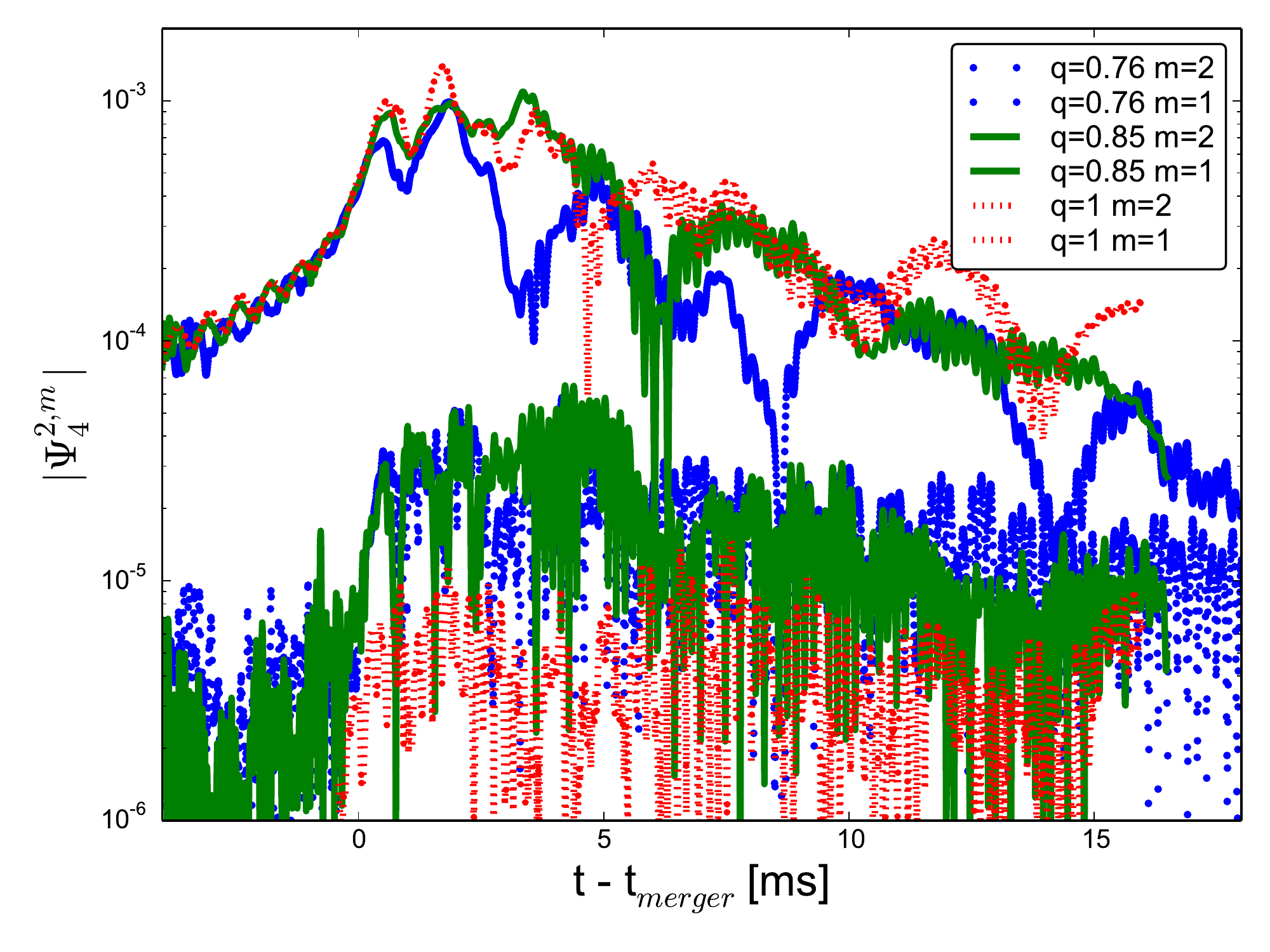}
\caption{The norm of a given mode $(l=2,m)$ of the gravitational radiation described by $\Psi^0_4$ as a function of time for different mass ratios with the DD2 EoS. 
Overall, the $(l=2,m=1)$ mode achieves saturation earlier for unequal mass than equal mass binaries.
Notice that the $m=1$ modes decay more slowly with time than the corresponding $(l=2,m=2)$ modes. 
Here, the merger time is chosen when the stars first come into contact.
}
\label{fig:modes_psi4}
\end{figure}

As discussed elsewhere (see e.g. Ref.~\cite{Lehner:2014asa} and references cited therein), the merger of neutron stars involves a violent collision
in which the individual stars move at a large fraction of the speed of light ($v \approx 0.3c$).
The newly formed massive neutron star rotates differentially with a primarily
quadrupolar structure that produces gravitational waves in the $l=2,m=2$ mode.
However, the one-armed spiral instability can develop so that the gravitational radiation includes an
$l=2,m=1$ component.
The development of this mode is apparent in Fig.~\ref{fig:modes_psi4} which displays 
the strength of the two $l=2$ modes (as measured by the Newman-Penrose radiative scalar $\Psi_4$) 
with the DD2 EoS for the three  mass ratios considered here.

The one-armed instability leads to the rapid rise of  the $m=1$ mode on a short time scale: $\lesssim 1.5$\,ms for the unequal mass
cases and $\lesssim 3$\,ms for the equal mass case. The equal mass case demonstrates a longer timescale as a
 result of the intrinsic symmetry of non-spinning,
zero-eccentricity, equal mass mergers.
Indeed, for the instability to occur in the equal mass case, a mechanism is required to break the symmetry to allow for odd modes. 
Such a mechanism is naturally provided by the Kevin-Helmholtz instability that arises in the contact 
region and in the associated  turbulence~\cite{Radice:2016gym,B_fields_are_us}; as well
a type of  magnetic (Tayler) instability can induce an $m=1$ perturbation~\cite{1999A&A...349..189S,B_fields_are_us}.

\begin{figure}[h]
\centering
\includegraphics[width=8.5cm,angle=0]{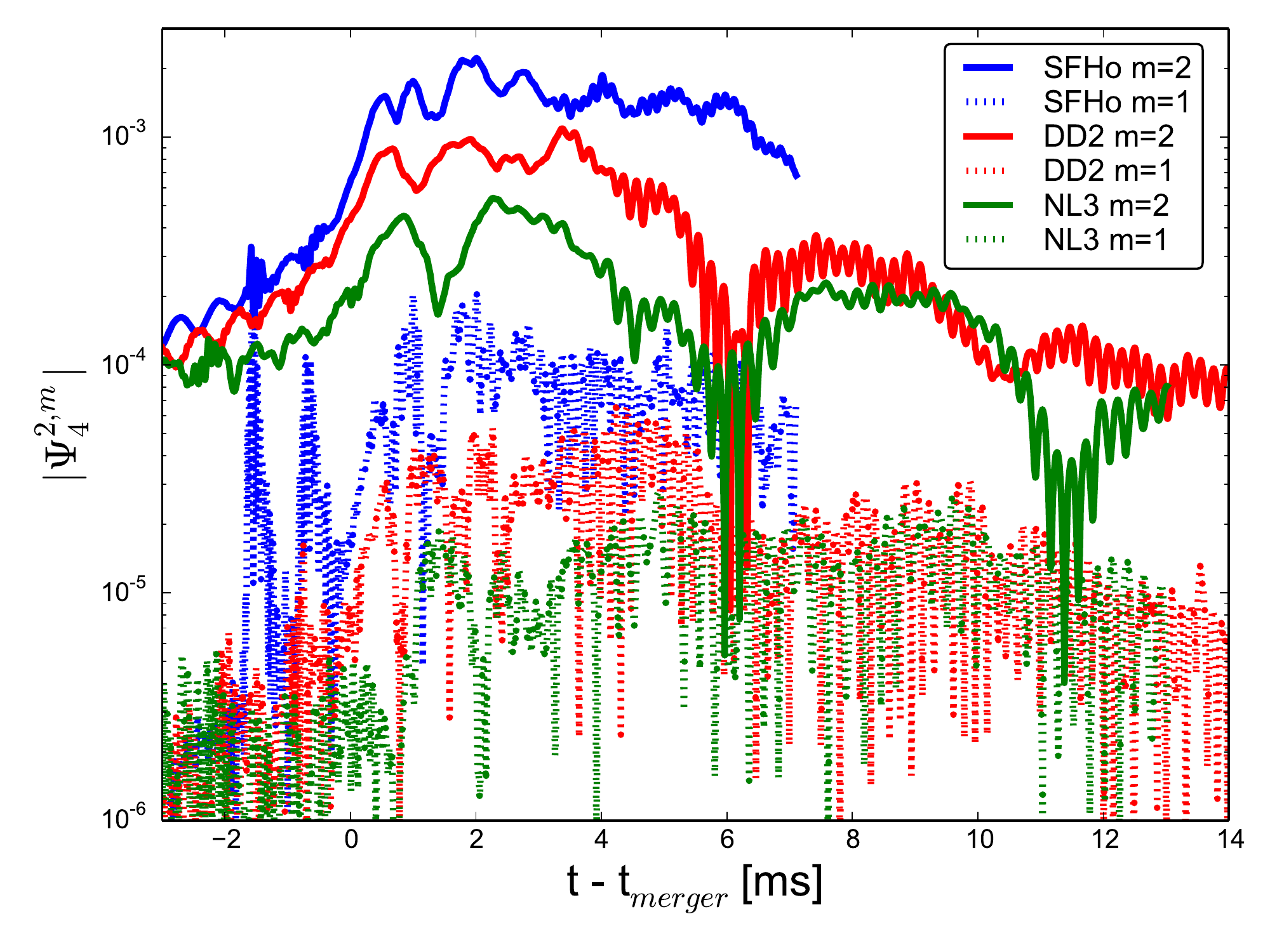}
\caption{ Same as Fig.~\ref{fig:modes_psi4} for q=0.85 with three different EoS (SFHo, DD2, NL3).
Although the qualitative behavior is similar for all the cases, the strength of the signal is stronger for the softest EoS. Additionally,  the decay rate of the $m=1$ mode has a mild dependence on the EoS though this
observation might be affected by
other physics effects---such as the Tayler instability.
Notice that the remnant with the SFHo EoS collapses to a black hole roughly $8$ms after merger.
}
\label{fig:modes_psi4_q085}
\end{figure}

Naturally, as the mass ratio departs from equality the $m=1$ mode becomes stronger and saturates
earlier.
This qualitative behavior generally holds among all three EoS,
as is apparent
in Fig.~\ref{fig:modes_psi4_q085} that presents both the $m=2$ and $m=1$ modes for $q=0.85$ 
for the three different EoS. In all three cases, the $m=1$ reaches its saturation value roughly $\approx 1.5$\, ms
after the stars come into contact.

This observed behavior is driven by the dynamics of the newly formed massive neutron star. 
Its fate will depend on the detailed structure of the object and, 
for example, its degree of differential rotation and
continued susceptibility to instabilities. One useful avenue to begin the 
analysis of the global structure of the resulting hypermassive star is to 
construct cylindrical profiles of the star and decompose these profiles
into Fourier amplitudes as was computed in Ref.~\cite{Ou:2006yd}, for example.

\begin{figure}[h]
\centering
\includegraphics[width=8.5cm,angle=0]{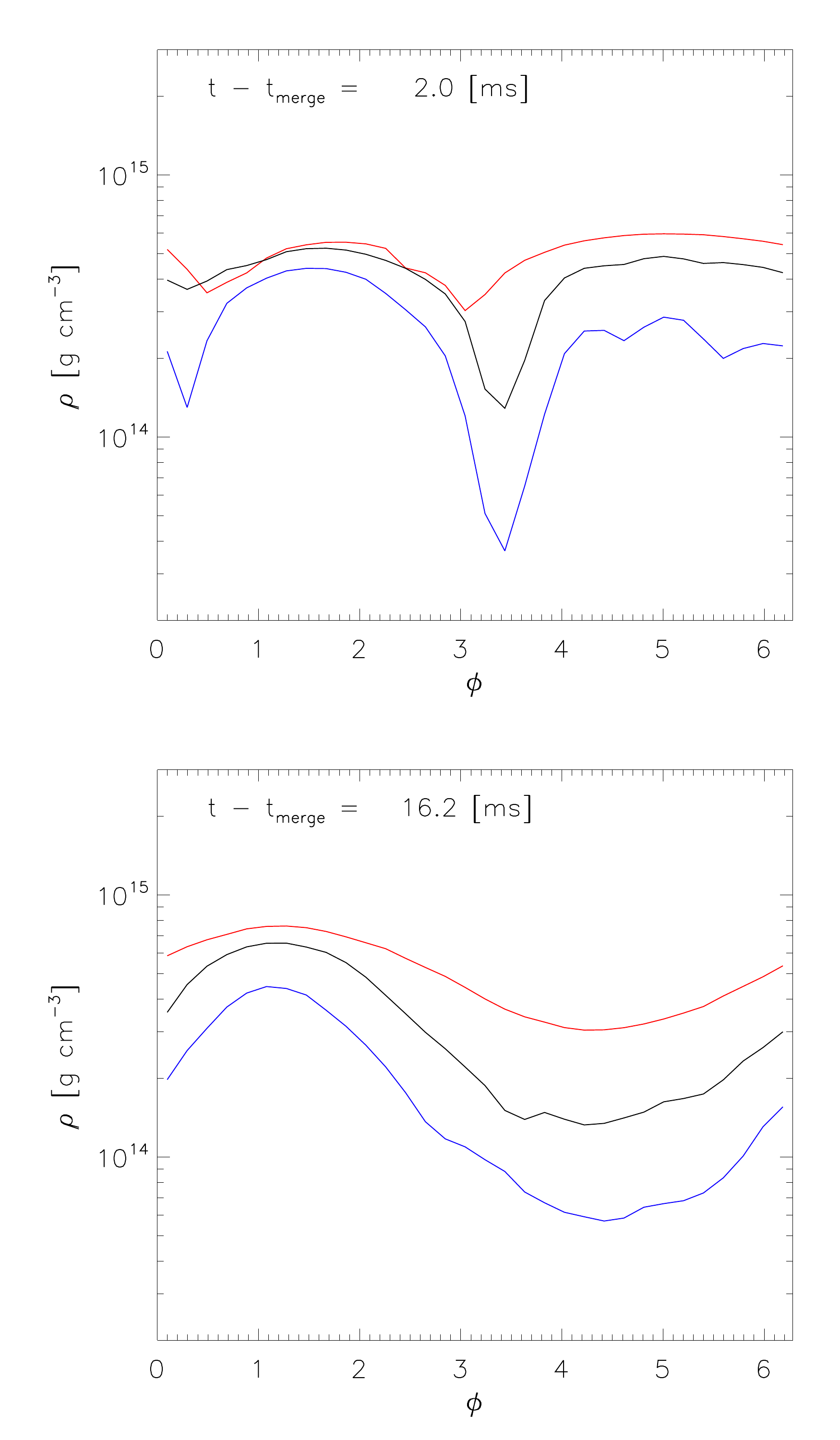}
\caption{The panels show
three profiles of the average mass density in the equatorial plane, 
extracted from annuli of width $290\mathrm{m}$  centered on the 
origin at radii of $4.4\mathrm{km}$ (red curve), $7.4\mathrm{km}$
(black curve), and $10.3\mathrm{km}$ (blue curve) from the 
$q = 0.85$ simulation with DD2 EoS. 
The density profiles $\approx 2$ms after merger (\textbf{top panel}) indicate a strong, $m = 2$ bar mode modulation;
at  $\approx 16$ms (\textbf{bottom panel})  on the other hand, the density variations a
are dominated by a single $m = 1$ mode.  The eigenmode can be seen to have
a trailing spiral character as the peak in the profile shifts to slightly 
smaller azimuth as the radius increases.
}
\label{fig:rho_prof}
\end{figure}

As an illustration of the density behavior and its qualitative change
through time, Fig.~\ref{fig:rho_prof} shows three such profiles extracted from 
the binary neutron star evolution with $q = 0.85$ using the DD2 
EoS. These profiles are extracted in three annuli in the equatorial
plane, each with a width of $290 \; \mathrm{m}$ and centered 
on radial distances of: $4.4 \; \mathrm{km}$ (red curve),
$7.4 \; \mathrm{km} $ (black curve), and 
$10.3 \; \mathrm{km}$ (blue curve) at times of 
$2.0$ and $16.2 \; \mathrm{ms}$ after merger in this system. 
The profiles use annuli that are evenly spaced in radius and use
32 evenly spaced azimuthal zones. A nearest cell algorithm is used
to construct the average mass density profile.
At the later point in the simulation, the remnant's structure is clearly
dominated by a $m = 1$ mode as the density profiles are well-matched
by a single sine wave, though there is also power at higher $m$ values
as well, particularly at larger radii. One can also note that the maximum
value in the profile shifts to slightly smaller azimuthal angle at increasing
radius---indicative of the trailing, one armed spiral wave of a
$m = 1$ mode.

\begin{figure}[tb]
\centering
   \includegraphics[scale=0.40]{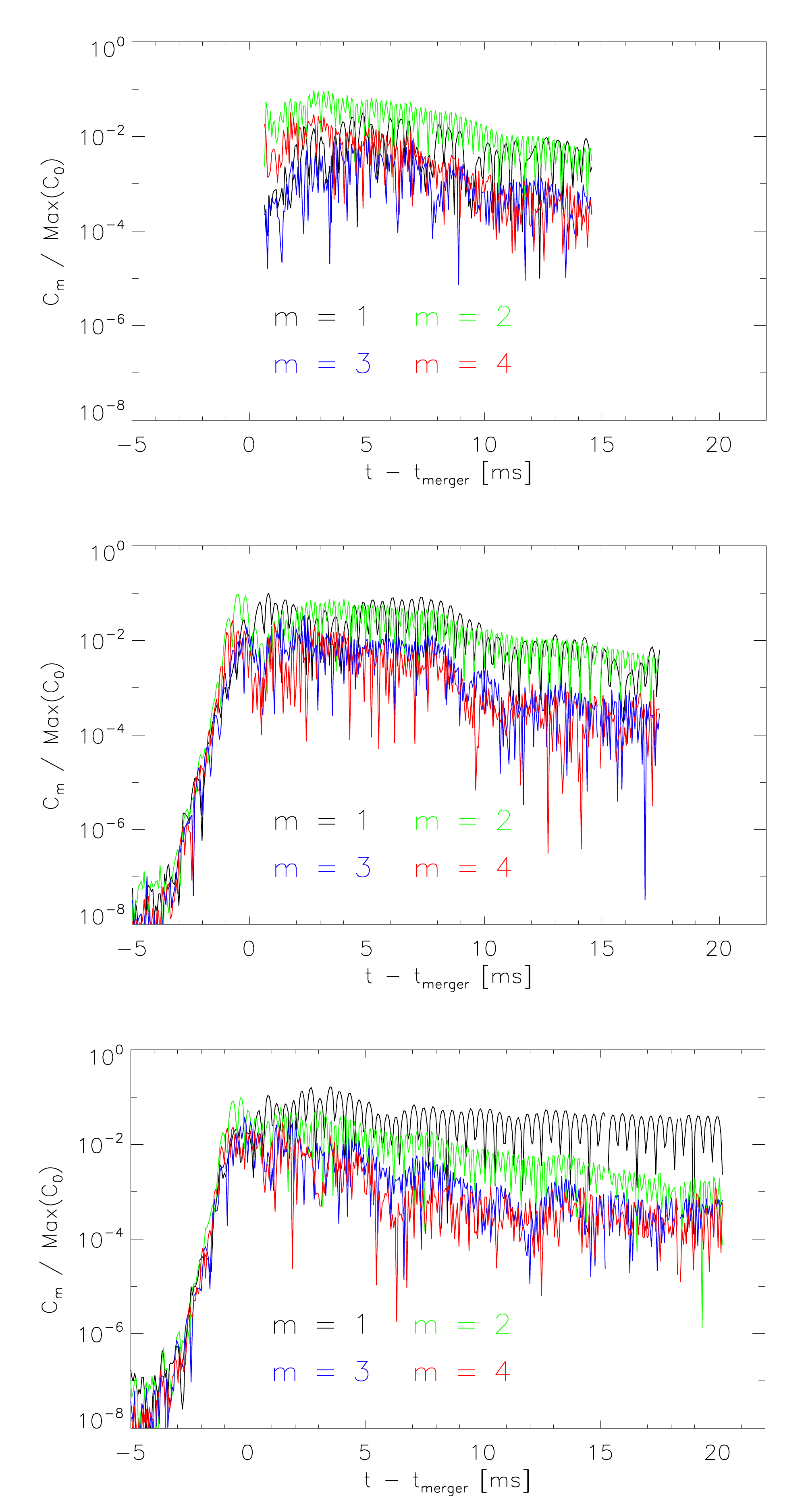}
\caption{Fourier decomposition of rest mass density extracted from an annulus
at $3.3 \; \mathrm{km} $ and $370 \; \mathrm{m} $ wide about the instantaneous
center of mass
from simulations of a symmetric binary (\textbf{top panel}), a binary with mass ratio
of $q = 0.85$ (\textbf{middle panel}) and $q = 0.76$ (\textbf{bottom panel}). These three simulations 
were evolved with the DD2 EoS. The amplitude of modes with $m$ from 1 to 4
are shown normalized by the largest value of the 0th component of the 
Fourier decomposition in the annulus.
}
\label{fig:rho_modes_ft}
\end{figure}

To help quantify this impression we perform a Fourier decomposition 
across the equatorial density distribution. We extract average density 
profiles in 10 evenly spaced annuli ranging in distance from 
$0.8 \; \mathrm{km}$ to $4.8 \; \mathrm{km} $ from the instantaneous 
center of mass
with a width of $370 \; \mathrm{m} $ across 32 evenly
spaced azimuthal zones. These profiles are then transformed to extract
complex amplitudes, $C_{m}$. The time evolution of these Fourier 
coefficients shows similar behavior across the 10 annuli and, for 
simplicity, we plot only the coefficients for one annulus in the 
middle of the set at a radius of $3.3 \; \mathrm{km} $
in Fig.~\ref{fig:rho_modes_ft}. 
The coefficients are normalized by the maximum value of the 
DC (m = 0) coefficient in the annulus over the course of the
simulation.

The middle panel of Fig.~\ref{fig:rho_modes_ft} shows Fourier
coefficients for the simulation with $q = 0.85$ initially. After 
merger, the $m = 1$ begins to dominate, as already indicated in the 
plot of the density profiles in Fig.~\ref{fig:rho_prof}.
The $m = 2$ mode decays indicating that the hypermassive star
does not form a bar (again, confirming the profiles in Fig.~\ref{fig:rho_prof})
 while both the $m = 3$ and the eventually numerically dominated
$m = 4$ mode show similar decaying behavior (at a slightly higher rate than the $m=2$ mode).
The bottom panel shows the analysis for the $q=0.76$ case, which are similar
to that for the $q=0.85$ case.
For the symmetric binary, shown in the top plot of 
Fig.~\ref{fig:rho_modes_ft}, the $m = 2$ bar mode dominates the post-merger evolution
for $\approx 15$ms, though an $m = 1$ has comparable power at the end of the simulation.
This behavior is expected because it takes time for the turbulence to break
the symmetry allowing the $m=1$ to develop. Hence, at first contact, the $m=1$ is essentially zero. 
In contrast, an $m=1$ component is already present at first
contact due to the asymmetry of the unequal mass case.
Nevertheless, as mentioned, after about a few ($\approx 5$)
rotation periods of the merged object the mode has saturated, and it overtakes
the $m=2$ mode in roughly $20$ rotation periods.

\begin{figure}[h]
\centering
\includegraphics[width=8.5cm,angle=0]{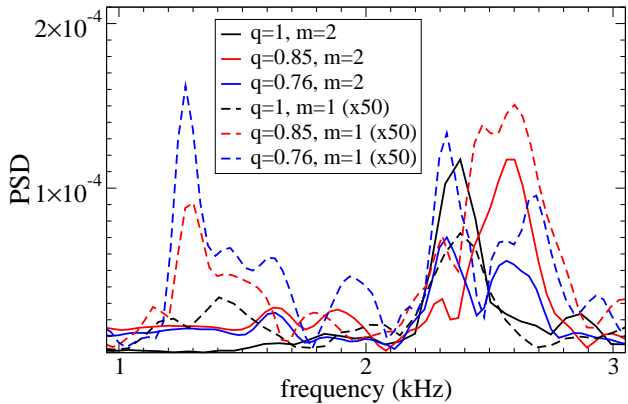}
\caption{Fourier transform of the $\Psi^0_4$-modes $(l=2,m)$ shown in Fig.~\ref{fig:modes_psi4} for the DD2 EoS. For this EoS, the dominant radiative mode after merger is the $(l=2,m=2)$ mode, peaking at a frequency $f_{\rm peak} = 2.3-2.6$~kHz. Notice that, in contrast, the $(l=2,m=1)$ mode peaks around $f_\mathrm{m1} \approx f_{\rm peak}/2 \approx 1.3$~kHz. This peak becomes more prominent for small mass ratios. 
}
\label{fig:modes_ft}
\end{figure}

Interestingly, the frequency of the $m=1$ gravitational wave mode appears insensitive to the mass ratio, $q$,
but is dependent on the EoS.
We obtain the $m=1$ frequencies from Fourier decompositions $\Psi_4$ (as shown
in Fig.~\ref{fig:modes_ft} for the DD2 case), and obtain
$f_\mathrm{m1} \approx 1.0, 1.25, 1.7$\,kHz for NL3, DD2, and SFHo, respectively.
Importantly these frequencies are half that of the dominant $m=2$ mode for the same EoS 
($f_{\mathrm{peak}} \approx 2,2.5,3.3$kHz) and therefore fall in a 
higher sensitivity band, albeit with (initially) a lower strain.  Furthermore, 
the measured values for $f_\mathrm{m1}$ agree with the rotational frequencies 
of the remnant densities.

Before examining the detectability, we comment on the decay rates of the $m=1$ mode.
In particular, while initially stronger than the $m=1$ mode, the $m=2$  decays at a faster
rate. We estimate the decay rate to be $\approx t^{-4\pm0.5}$ for the $m=2$ mode of $|\Psi_4|$ 
while for the $m=1$ mode $\approx t^{-1.5 \pm 1}$. These exponents depend mildly
on the mass ratio $q$, as apparent in Fig.~\ref{fig:modes_psi4_q085}. 
A precise estimate of the rates is beyond the scope
of this note because of their dependence on a variety of physical effects such as  EoS, magnetohydrodynamics 
(requiring very high numerical resolution),
and, depending
on the lifetime of the remnant, transport effects that might become significant.
Nevertheless, the $m=1$ mode is clearly more weakly damped 
than the $m=2$, indicating the latter will overtake it 
(see also the discussion in~\cite{Radice:2016gym,East:2015vix}).

\section{Detectability}
As described above, the $m=1$ mode grows rapidly and saturates with a mostly
constant frequency and with an amplitude that decays slowly. 
For sufficiently massive systems, this mode ends  when
a black hole forms. For cases that do not promptly collapse, the lifetime
can easily be $T \gtrsim 20$\,ms. A precise value for $T$ depends on
the details of the system such as
the total mass and the particular EoS. Also important are the dynamics
of the remnant that could transport angular momentum outward and reduce thermal
pressure support via cooling.
Although an estimate of the neutrino cooling is considered in these simulations with a leakage scheme, we are not accounting for magnetic and radiation transport effects, which could
potentially affect the $m=1$ mode over timescales of order $\approx 100$\,ms.
Thus, in the following we adopt a conservative baseline lifetime of $T =10$\,ms.

To estimate the SNR
that such a mode would produce, we make use of the analysis
described in Ref.~\cite{Jaranowski:1998qm}. The SNR $\rho$ of a monochromatic gravitational
wave with strain $h=|\Psi_4^0|/(L M_{\odot} \omega^2)$  [recall, asymptotically $\Psi_4 = \Psi_4^0/r + {\cal O}(r^{-2})$], at a fixed gravitational wave frequency $f=\omega/(2\pi)$ (from a source at
a distance $L$) in a time window $[T_i, T_f]$,
can be estimated by
\begin{equation}
\rho^2 \simeq \frac{2}{S_n(f)} \int_{T_i}^{T_f} h^2\,dt\, . \label{snr1}
\end{equation}
The noise strain of the detector is denoted by $\sqrt{S_n(f)}$ and given in units of $\sqrt{\mathrm{Hz}^{-1}}$.

In all our examined cases corresponding to the DD2 EoS, the frequencies $f_\mathrm{m1}$ of $m=1$ modes agree with
those measured from the orbital rotation rates. 
Additionally, we will assume a typical value $|\Psi^0_{4_{m=1}}| \approx 5 \times 10^{-5}$ as observed in our simulations.
With the exception of the SFHo case which collapses to a black hole roughly $8$\,ms after merger,
the other cases result in a long-lived remnant that we have evolved for at least $15$\,ms post-merger~(Ref.~\cite{Sekiguchi:2016bjd} has followed the equal-mass DD2 case for $\approx 30$\,ms after merger).

Using  Eq.~\ref{snr1} and {\em assuming a constant strain}, we arrive at an estimate of the 
SNR for the $m=1$ mode for a source at a horizon distance $L$ as,
\begin{eqnarray}
\rho_{m=1} & \approx & 11 \times
                 \left[ \frac{6\,\times 10^{-24} \mathrm{Hz}^{-1/2}}
                        {\sqrt{S_n(f_\mathrm{m1})}} \right]
                 \left[ \frac{|\Psi^0_{4_{m=1}}|}{5 \times 10^{-5}} \right]   \nonumber \\
     &         & 
                 \left[ \frac{1.3 \mathrm{kHz}}{f_\mathrm{m1}} \right]^{2}
                 \left[\frac{T}{10\mathrm{ms}}\right]^{1/2}
                 \left[ \frac{10 \mathrm{Mpc}} {L} \right]
\label{snr}
\end{eqnarray}
where we have used for aLIGO $\sqrt{S_n(f=1.3 \mathrm{kHz})} \approx 6 \times 10^{-24} \mathrm{Hz}^{-1/2}$
(as per the zero-detuned, high laser power (no signal-recycling mirror) noise curve estimated 
in~\cite{LIGOCURVE}). 

Longer times will increase the SNR, and for scenarios with
constant strain, the SNR would increase as $\sqrt{T/10\mathrm{ms}}$. The evolutions of Ref.~\cite{Radice:2016gym}
suggested that the strain can indeed be fairly constant for 
resolutions equal or better than
$\Delta x \approx 222$m (while they see decay for $\Delta x = 295$m). Our resolution, $\Delta x=230$m, is 
comparable though it could be possible that the rates of decay measured would decrease
further with higher resolution. We caution however that for long-time scales other physics will
come into play---for example cooling and angular momentum transfer---thus at this point it is difficult to draw
firm conclusions on timescales longer than $\approx 20$ms. At a conservative level, however,
we can employ the decay rate we observe here for the $m=1$ mode, $\propto t^{-1.5}$, and estimate the SNR 
for $t\in[3,20]$\,ms as
\begin{eqnarray}
\rho_{m=1} & \approx & 2 \times
                 \left[ \frac{6\,\times 10^{-24} \mathrm{Hz}^{-1/2}}
                        {\sqrt{S_n(f_\mathrm{m1})}} \right]
                 \left[ \frac{|\Psi^0_{4_{m=1}}(3\mbox{ms})|}{5 \times 10^{-5}} \right]   \nonumber \\
     &         & 
                 \left[ \frac{1.3 \mathrm{kHz}}{f_\mathrm{m1}} \right]^{2}
                 \left[ \frac{10 \mathrm{Mpc}} {L} \right].
\label{snr2}
\end{eqnarray} 
This value is certainly lower (at a given distance) than that assuming a constant magnitude in 
Eq.~(\ref{snr}) but
is  nevertheless significant. 

As a figure of merit, we can compare the SNR of the $m=1$ mode to the $m=2$ mode using
the same assumptions. In particular, assuming the $m=1$ and $m=2$ modes emit at a constant
frequency and with decay rates given by $\propto \{t^{-1.5},t^{-4}\} $ respectively, and
also making use of the fact that the noise curve of aLIGO grows approximately 
linearly with increasing frequency. The ratio of the corresponding SNRs in a time window 
$[T_i, T_f]$, results in
\begin{eqnarray}
\frac{\rho_{m=1}}{\rho_{m=2}} &=& 
\frac{\sqrt{S_n(2 f_0)}}{\sqrt{S_n(f_0)}} \frac{|h_{m=1}(T_i)|}{|h_{m=2}(T_i)|} 
\sqrt{\frac{14}{4}}\, \sqrt{\frac{(1-(T_i/T_f)^2)}{(1-(T_i/T_f)^7)}} \nonumber \\
&\approx& 13 \frac{|\Psi^0_{4_{m=1}}(T_i)|}{|\Psi^0_{4_{m=2}}(T_i)|} .
\end{eqnarray}
Therefore, if  at $t=T_i$, the $m=1$ mode strength of $\Psi_4^0$ is at least a thirteenth of the strength of the $m=2$ mode, 
it will become more relevant than the $m=2$ as a contributor to the post-merger gravitational wave
signal (a condition that would be even less strict if the $m=1$ mode decays at a slower rate).

The estimate obtained in Eq.~\ref{snr} is not, at
first sight, encouraging for possible 
detection in a single event. 
However, close binaries with a very long lived $m=1$ mode would be the most promising for this task,
and of course future detectors and upgrades of current detectors will naturally improve prospects for detecting this
post-merger signal. However, we argue here that approaching the detection of the $m=1$ mode using {\em triggered search strategies}
significantly enhances the prospects. The detection of the inspiral signal from a distant
neutron star binary would provide timing estimates and physical parameters. In particular
knowledge of the total mass, in turn, will inform whether a long lived remnant is probable. This
information would allow for the same sort of aggressive data analysis 
and reduced detection thresholds as occur in follow-ups of electromagnetic
triggers~\cite{Abbott:2008mr,Dietz:2012qp,TheLIGOScientific:2013cya}. 
The inspiral would serve as the trigger for both the post-merger $m=1$ and $m=2$ analysis, 
which can be correlated, and this trigger would allow for a decreased threshold in SNR for detection.
In particular assuming a detection threshold in SNR of $\approx 1$, then Eq.~\ref{snr} yields
a horizon distance as high as $100\,\mathrm{Mpc}\left[\sqrt{T/10\mathrm{ms}}\right]$ 
(again, assuming that the mode maintains a constant strain for that time period).

The detection of both modes will provide information about the EoS. The observation
that the $m=1$ mode frequency is half that of the $m=2$ mode means that the peak frequencies
of both modes encode information about the EoS as has been demonstrated for the $m=2$
modes in recent work~\cite{2015arXiv150203176B,2015PhRvL.115i1101B,Lehner:2016lxy}.
For instance, Ref.~\cite{Lehner:2016lxy} finds a fit for this peak frequency
\begin{equation}
f_{\rm peak} [\mathrm{kHz}] = -1.61 + 2.96 f_{\rm c}\left[ \frac{2.7 M_\odot}{M} \right] [\mathrm{kHz}] \label{fitestimate}
\end{equation}
in terms of the ``contact'' frequency $f_{\rm c}$. This frequency is strongly dependent on the EoS and is defined in terms of the gravitational masses 
involved, $m_g^{(1)}$ ($M_g=m_g^{(1)}+m_g^{(2)}$) and their compaction ratio $C_i$ 
\begin{equation}
  f_{\rm c} = \frac{1}{\pi M_g} \left( \frac{m_g^{(1)}}{M_g C_1} + \frac{m_g^{(2)}}{M_g C_2}  \right)^{-3/2} .
\label{fcontact}
\end{equation}
Data analysis can be directed to look for a correlated signal in these two modes that may
improve the extraction of the encoded EoS information.

\section{Conclusions}
\label{sec:conclusions}
The evolutions presented here, as well as those in Refs.~\cite{Radice:2016gym,East:2015vix}, 
indicate that the one-armed spiral instability develops generically in the merger of binary
neutron stars (that do not promptly collapse to a black hole). Our analysis includes both (non spinning) equal
and unequal mass binaries described by realistic equations of state. We find that this instability
develops for a broad range of EoS ranging from soft to stiff, though its strength is weaker for
stiff EoS than for soft ones.
For the equal mass cases, the instability is seeded by turbulence while in the unequal mass cases
the $m=1$ mode is seeded strongly from the onset because of the inherent asymmetry of the system. 
In agreement with Refs.~\cite{Radice:2016gym,East:2015vix} we find that the mode grows quickly and saturates in
just a few rotational periods of the newly formed object. 
For unequal mass mergers within $\lesssim 14 \mathrm{Mpc} \left[ \sqrt{T/10\mathrm{ms}}\right]$,
the strength of the mode is
sufficient for a {\em direct detection} (i.e. with SNR$\ge 8$) by aLIGO with planned sensitivities (in particular
with the so-called ``zero-detuned, high laser power (no signal-recycling mirror)'' configuration) provided
the mode stays roughly constant in strength. If this mode decays as estimated in this work
$\propto t^{-1.5}$, the horizon distance for detection is reduced to $\simeq 3$Mpc though this
decay could be reduced in higher resolution studies (see~\cite{Radice:2016gym}). 

The time for the $m=1$ mode to reach saturation naturally depends
on the asymmetry of the merger which, for non-spinning binary neutron stars, is determined by
the mass ratio of the binary. Even in equal mass, symmetric encounters this time is observed to be at most $4$\,ms.
Although the strain of the $m=1$ mode in all cases roughly $4$\,ms after merger is weaker 
than the dominant, $m=2$ mode, its strain can be as large as $\approx 50\%$ of the main $l=2,m=2$ mode. 
Furthermore the ratio of strain in the $m=1$ to $m=2$ modes grows with time and becomes larger than unity
after $\approx 20-30$ms (for systems that do not collapse to a black hole within this timeframe).
The $m=1$ mode also benefits from having a frequency half that of the $m=2$ mode where
aLIGO has less noise (the noise curve grows roughly quadratically in frequency in this region frequency region).
Because of the slower decay time and smaller frequency, the SNRs of both modes become roughly
comparable and that of the $m=1$ should dominate at late enough times.

We have also presented 
a more hopeful prospect for detection than in Ref.~\cite{Radice:2016gym}.
Because gravitational waves from the inspiral of a binary neutron
star system would be detectable by aLIGO/VIRGO at distances of a few $100s$ of megaparsecs, the search for the post-merger
$m=1,2$ modes would proceed as in a triggered-search. 
In particular, because we have a good estimate both of the mode frequencies and of their timescales, 
such an analysis would be quite well targeted.
A Bayesian analysis 
may be able to
extract characteristic features from the post-merger signal as long as differences of ${\cal O}(1)$ arise
from the assumption of various priors motivated by the dynamics discussed here.
Thus a targeted analysis may significantly extend the reach of the modest distance assumed in Eq.~\ref{snr}.
Finally, the {\tt 2:1} connection of the peak frequencies of these two modes with the EoS 
could enable stacking different signals and enhance the possibility of
extracting these modes.

The detection of these modes would complement a detection of the pre-merger inspiral, providing more information
about the EoS. As recent work has shown, the post-merger remnant demonstrates a characteristic frequency peak
that is intimately related to the underlying EoS~\cite{2015arXiv150203176B,Dietrich:2015iva,Bauswein:2015vxa,2015PhRvL.115i1101B,Lehner:2016lxy,Rezzolla:2016nxn}.

We have studied the $m=1$ spiral arm instability 
that, along with the $m=2$ mode,  encodes important information about the 
post-merger remnant and its EoS.  The dynamics of the $m=1$ mode are important for
its ultimate detection by both current and future detectors, and it is therefore
important  to better understand the impact of various effects 
including cooling, transport of angular momentum, and magnetic
instabilities. Further work on these fronts is ongoing~\cite{ericinpreparation}.

%
%
\vspace{0.75cm}
~\\

\begin{acknowledgments}
It is a pleasure to thank William East, Gabriela Gonzalez, Chad Hanna,
Sascha Husa, Francisco Jimenez, Vasileos Paschalidis, and
Frans Pretorius for interesting discussions as well as our 
collaborators Eric Hirschmann, David Neilsen, and Marcelo Ponce.
This work was supported by NSF grant PHY-1308621~(LIU),
NASA's ATP program through grant NNX13AH01G,
NSERC through a Discovery Grant (to LL) and CIFAR (to LL).
CP acknowledges support from the Spanish Ministry of Education and
Science through a Ramon y Cajal grant and from the Spanish Ministry of
Economy and Competitiveness grant FPA2013-41042-P.
Research at Perimeter Institute is supported through Industry Canada and 
by the Province of Ontario
through the Ministry of Research \& Innovation.  Computations were
performed at XSEDE and Scinet. 
\end{acknowledgments}

\bibliographystyle{utphys}

\providecommand{\href}[2]{#2}\begingroup\raggedright\endgroup
\end{document}